\documentclass[preprint,authoryear,12pt]{elsarticle}

\usepackage{amssymb}
\usepackage{amsmath}
\usepackage{amssymb}
\usepackage{natbib}
\usepackage{graphics}
\usepackage[latin1]{inputenc}
\usepackage[T1]{fontenc}
\usepackage{rotate}
\usepackage{color}
\usepackage{wrapfig}
\usepackage{latexsym,epsfig,fancyheadings}
\usepackage{latexsym,epsfig}
\graphicspath{{Figures/}}

\journal{International Journal of Mechanical Sciences}

\begin{document}

\begin{frontmatter}

\title{Solution methods for the growth of a repeating imperfection in the line of a strut on a nonlinear foundation}
\author{R. Lagrange\corref{cor1}}
\ead{romain.g.lagrange@gmail.com}
\ead[url]{http://rlagrange.perso.centrale-marseille.fr/visible/Site/}

\author{D. Averbuch\corref{cor2}}
\ead{daniel.averbuch@ifpen.fr}

\cortext[cor1]{Principal corresponding author}
\cortext[cor2]{Corresponding author. Phone: +33 437 702 000}

\title{Solution methods for the growth of a repeating imperfection in the line of a strut on a nonlinear foundation}

\address{IFP \'Energies nouvelles, Rond-point de l'\'echangeur de Solaize, BP3, 69360 Solaize, France}

\begin{abstract}
This paper is a theoretical and numerical study of the uniform growth of a repeating sinusoidal imperfection in the line of a strut on a nonlinear elastic Winkler type foundation. The imperfection is introduced by considering an
initially deformed shape which is a sine function with an half
wavelength. The restoring force is either a bi-linear or an
exponential profile. Periodic solutions of the equilibrium problem
are found using three different approaches: a semi-analytical
method, an explicit solution of a Galerkin method and a direct
numerical resolution. These methods are found in very good agreement
and show the existence of a maximum imperfection size which leads to
a limit point in the equilibrium curve of the system. The existence
of this limit point is very important since it governs the
appearance of localization phenomena.

Using the Galerkin method, we then establish an exact formula for
this maximum imperfection size and we show that it does not depend
on the choice of the restoring force. We also show that this method
provides a better estimate with respect to previous publications.
The decrease of the maximum compressive force supported by the beam
as a function of the imperfection magnitude is also determined. We
show that the leading term of the development has a different
exponent than in subcritical buckling of elastic systems, and that
the exponent values depend on the choice of the restoring force.

\end{abstract}


\end{frontmatter}

\section{Introduction}
The subject of beam buckling can be found in several situations in
industrial applications. Among the most studied are the thermal
track buckling and the buckling of subsea pipelines under the effect
of temperature and/or pressure. In the latter case, buckling can
appear and in both the vertical and horizontal planes, according to
the existing restraints imposed by the environment and the backfill.
Numerous authors have studied these two applications due to their
high practical importance, and have proposed solution methods to
determine buckling loads and post-buckling situations. Along time,
the techniques used have progressed, based firstly on analytical
analyses and latter on numerical methods mostly derived from finite
elements models. Thus, based on some early work by
\cite{Kerr1,Kerr2} studying the stability of railway tracks
subjected to thermal buckling, several authors such as
\cite{Bournazel3,Hobbs5,Hobbs4} have proposed solution methods where
the equilibrium equations were solved in post-buckling
configurations to establish relevant buckling loads. In these works,
the soil was supposed rigid, while the external forces acting on the
beam was assumed as constant as a dead weight or constant friction
force. One of the key features of these theories is the fact that
the loss of contact (or movement) induces a loss of global stiffness
of the structure which leads to subcritical buckling and infinite
buckling loads if no imperfection is assumed. Using slightly
different arguments, other models were proposed in \cite{Croll6} and
\cite{Maltby8}. In the latter work, equilibrium equations were
obtained by assuming sine deflections in the post-buckling
situations and using an approximate Galerkin solution method. This
method was compared with numerical solutions in \cite{Maltby7} and
against experimental results in \cite{Maltby8}. Though using an
approximate solution method, the approach showed good results with
respect to the numerical simulation. In order to improve the earlier
methods, numerical models were developed in
\cite{Ju9,Klever11,Leroy12,Yun10} to incorporate for instance
additional non linear effects in the models, such as non linear
geometric and material models. One of the key aspects of work
related to the study of the upheaval buckling is the study of the
localization phenomenon, which was suspected early in the pioneering
work of \cite{Tveergard13,Tveergard14}. This aspect was analyzed
through numerical simulations in these first papers and in
\cite{Hunt16,HuntBlack}, and through analytical approaches based on
a double scale expansion of the equilibrium equations in
\cite{Potier15}, for a beam resting on an elastic non-linear
foundation. Based on these results, the study of the localization
phenomenon, was continued by using Galerkin techniques
\cite[see][]{Wadee18,Whiting17} using the displacement envelopes
obtained through the double-scale expansion of \cite{Potier15}. A
lot of attention has also been paid to the estimation of the
mechanical restraint induced by the soil friction and to the effect
of backfill on the pipeline behavior, since the corresponding forces
were found to highly influence the mechanical behavior of the pipe.
In order to feed the corresponding models, experiments were
performed \cite[see][]{Palmer20,Schaminee19,Trautmann22,Trautmann21}
either through centrifuge testing of small-scale pipeline models or
through direct testing of buried full scale pipe sections.

The present paper is an attempt to provide additional solutions for the
study of the growth of a repeating imperfection in the line of a strut on a nonlinear foundation. In
this work, the foundation is supposed to act through an either
bi-linear or exponential regularized friction model relating the
interaction line force to the transverse displacement (see section
\ref{Section2}). These two kinds of models are indeed found in the
above mentioned papers describing the soil-pipe interaction models.
In the former case, a solution method (piecewise solution) in
section \ref{Section3} is proposed by explicitly solving the
equilibrium equation in the regions where the foundation acts
linearly and where the friction force is constant, and by connecting
the two solutions by adequate boundary solutions. Alternatively, a
Galerkin approach of the same problem is developed in section
\ref{Section3} and leads to an explicit solution of the problem,
which is developed for the two regularization models. The piecewise
solution and the Galerkin approach are consequently compared
together and with numerical solutions of the problem in section
\ref{Section4}. The post-buckling problem is then studied through the Galerkin approach which
provides precise analytical solutions, focusing on the
characteristics of a limit point (see section
\ref{sectionPointLimite}) in the equilibrium curve depending on the
magnitude of the initial imperfection.

\section{Formulation of the differential equation}\label{Section2}
\begin{figure}
\begin{center}
\includegraphics[width=0.7\textwidth]{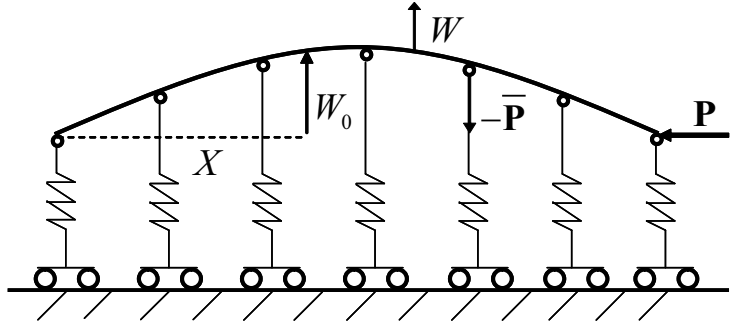}
\end{center}
\caption{Half-wavelength imperfection in the line of a strut resting on an elastic foundation. The
imperfection is $W_0$, the compressive load is $P$ and the buckling
displacement is $W$. Springs have a nonlinear elastic
force-displacement relationship $\overline
P\left(W\right)$.}\label{ProblemeModele}
\end{figure}

This section formulates the differential equation for the growth of an imperfection in the line of a strut on a nonlinear
Winkler-type foundation, see figure \ref{ProblemeModele}.
The imperfection is introduced by supposing an initially
deformed shape $W_0$ whose form is

\begin{equation}
W_0=A_0\sin \left(\frac{\pi }{L} X\right),
\end{equation}
with $A_0$ the amplitude of the imperfection, $L$ its length and $X$ the longitudinal coordinate. The compressive load is
$P$ and the restoring force per unit length is $\overline{P}$. These
two forces are assumed to be conservative. The differential equation
governing the deflection $W$ may be derived either: directly by
equilibration of forces; or from the Principal of Virtual Work; or
using an energy formulation. The latter approach is adopted here;
the total potential energy at first order being

\begin{equation}\label{EnergiePotentielle}
V = \int\limits_0^L {\left[ {\frac{1}{2}EI{W^{''}} ^2  -
P\left({\frac{1}{2}{W^{'}}^2 +{{W_0}^{'}}{{W}^{'}} }\right)  -
\int\limits_0^W {\overline P \left( t \right)dt} }
\right]{\rm{dX}}},
\end{equation}
where a prime indicates differentiation with respect to $X$. The
first term is the strain energy of bending ($EI$ is the bending
stiffness of the strut), the second is the work done by the load $P$
and the remainder is the energy stored in the elastic foundation.
Equilibrium is given by stationary values of $V$. In what follows
the strut is assumed to be simply supported, such that the
conditions at the boundaries are $W\left( 0 \right)= W\left( L
\right)=0$. The calculus of variations on (\ref{EnergiePotentielle})
gives, for a virtual displacement $\delta W$ such that $\delta
W\left( 0 \right)= \delta W\left( L \right)=0$

\begin{equation}\label{VariationEnergiePotentielle}
\delta V = EI{\left[ {\delta W^{'} W^{''} } \right]_0}^L  +
\int\limits_0^L {\left[ {EI{\rm{ }}W^{''''}  + P\left( {W^{''} + W_0
^{''} } \right) - \overline P \left( W \right)} \right]\delta
W{\rm{dX}}},
\end{equation}
so that the Euler-Lagrange equation and the conditions at the
boundaries for a simply supported strut are

\begin{subequations}\label{chap4_eq1Dim}
\begin{eqnarray}
EI W^{''''}  +P W^{''} -\overline P\left( W \right) &=&- P W_0 ^{''},\label{chap4_eq11Dim}\\
W\left( 0 \right) &=& 0,\label{chap4_eq12Dim}\\
{{ W''}}\left( 0 \right) &=& 0,\label{chap4_eq13Dim}\\
W\left( L \right) &=&0,\label{chap4_eq14Dim}\\
{{ W''}}\left( L \right) &=& 0.\label{chap4_eq15Dim}
\end{eqnarray}
\end{subequations}

In equation (\ref{chap4_eq11Dim}), $P$ is the compressive load
before buckling. The compressive load after buckling, considering a
strut of section $S$, should be written as $N=P-
\frac{{ES}}{{2L}}\int\limits_0^L {\left( {W_{,X} } \right)^2 dX}$,
last term of this expression being a geometric shortening which
allows for the additional length introduced by the lateral movement.
Therefore, $N$ should be used in the equation for equilibrium.
However, \cite{Tveergard14} have shown that the buckle will only
become unstable if $N(y)$ has a maximum is correct for an isolated
half-wave but is not correct for a long strut which contains a
sequence of half-waves end-to-end. In such a case the key point is
that a localization of the buckling, in which one particular
half-wave grows at the expenses of its neighbours, can occur whenever
the curve $P(y)$ has a maximum. Under this consideration, and as
\cite{Maltby8} did, we use $P$ instead of $N$ as the load parameter.

\subsection{The restoring force}
\begin{figure}
\begin{center}
\includegraphics[width=1\textwidth]{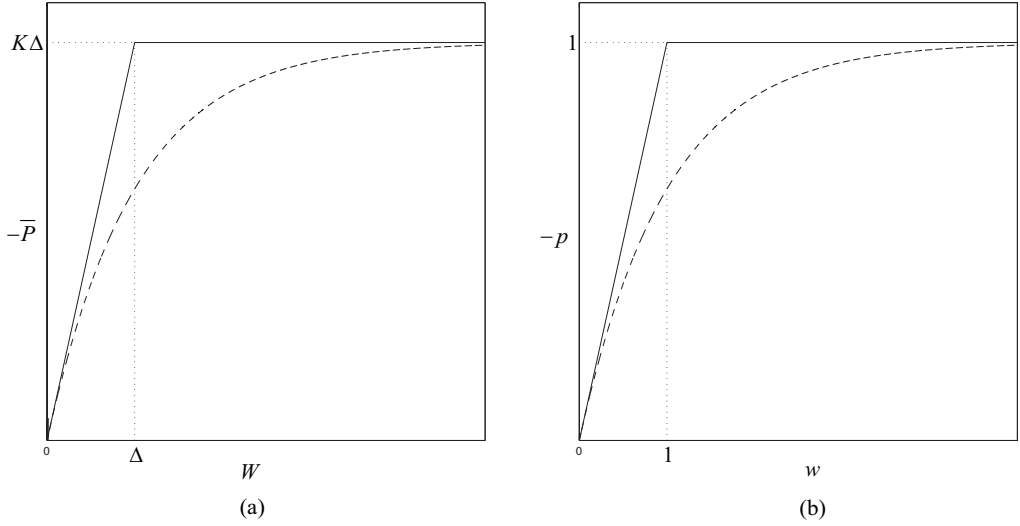}
\end{center}
\caption{(a) Restoring force. (b) Dimensionless restoring force.
Bi-linear profile (solid line), exponential profile (dashed line).
Dotted lines: limiting plateau (horizontal line), mobilization
(vertical line).}\label{Fig2_4}
\end{figure}

The restoring force per unit length is assumed to be nonlinear and
two particular $\overline{P}$ functions are considered. The first
one is referred as the bi-linear function and is defined by

\begin{eqnarray}
\overline{P}\left( W \right) = \left\{ \begin{array}{l}
  - KW\hspace{1cm}{\rm{ if}}\hspace{0.1cm}\left| W \right| < \Delta , \\
  - K\Delta \hspace{1.1cm}{\rm{ if }}\hspace{0.38cm}W > \Delta {\rm{ }}, \\
  K\Delta \hspace{1.40cm}{\rm{   if }}\hspace{0.38cm}W <  - \Delta,
 \end{array} \right.
\end{eqnarray}
where $K$ is the linear stiffness and $\Delta$ the mobilization.
These two constants are positive. The second $\overline{P}$ function
considered in this paper is referred as the exponential profile and
is defined by

\begin{eqnarray}
\overline{P}\left( W \right) = \left\{ \begin{array}{l}
  - K\Delta \left( {1 - e^{ - \frac{W}{\Delta }} }
\right) \hspace{0.85cm}{\rm{ if }}\hspace{0.38cm}W > 0{\rm{ }}, \\
   K\Delta \left( {1 - e^{ \frac{W}{\Delta }} }
\right)\hspace{1.40cm}{\rm{   if }}\hspace{0.38cm}W <0.
 \end{array} \right.
\end{eqnarray}
The two $\overline{P}$ functions share the same initial slope and
limiting force (see figure \ref{Fig2_4}(a)).

\subsection{Nondimensionalization}
Let's introduce a characteristic length $L_{char}  = \left(
{\frac{{EI}}{K}} \right)^{\frac{1}{4}}$ and nondimensional
quantities

\begin{eqnarray}
l = \frac{{ L }}{{L_{char} }}, x= \frac{{ X }}{{L_{char} }}, w =
\frac{{W}}{{\Delta}}, w_0 = \frac{{W_0}}{{\Delta}}, \lambda  =
\frac{P}{{K{L_{char}} ^2 }}, p=\frac{\overline{P}}{{K\Delta }}.
\end{eqnarray}
Hence, from (\ref{chap4_eq1Dim}) the deflection $w$ is solution of
the differential problem

\begin{subequations}\label{chap4_eq1}
\begin{eqnarray}
w^{''''}  + \lambda w^{''} -p\left( w \right) &=&- \lambda w_0 ^{''},\label{chap4_eq11}\\
w\left( 0 \right) &=& 0,\label{chap4_eq12}\\
{{ w''}}\left( 0 \right) &=& 0,\label{chap4_eq13}\\
w\left( l \right) &=&0,\label{chap4_eq14}\\
{{ w''}}\left( l \right) &=& 0,\label{chap4_eq15}
\end{eqnarray}
\end{subequations}
the dimensionless imperfection being
\begin{equation}\label{EquationDefautAdim}
w_0=a_0\sin \left(\frac{\pi }{l} x\right),
\end{equation}
with $a_0=\frac{A_0}{\Delta}$.

After nondimensionalization, the bi-linear function rewrites

\begin{eqnarray}
p\left( w \right)  = \left\{ \begin{array}{l}
  - w \hspace{1.35cm}{\rm{ if}}\hspace{0.1cm}\left| w \right| < 1, \\
  - 1\hspace{1.43cm}{\rm{ if }}\hspace{0.38cm}w > 1, \\
 1\hspace{1.75cm}{\rm{   if }}\hspace{0.38cm}w <  - 1,  \\
 \end{array} \right.
\end{eqnarray}
such that the dimensionless mobilization and limiting force equal 1.
The nondimensional exponential profile is given by

\begin{eqnarray}\label{LoiRessortRegularise}
p\left( w \right)= \left\{ \begin{array}{l}
  -  \left( {1 - e^{ - w} }
\right) \hspace{0.83cm}{\rm{ if }}\hspace{0.38cm}w > 0{\rm{ }}, \\
     {1 - e^{ w} }
\hspace{1.85cm}{\rm{   if }}\hspace{0.38cm}w <0.
 \end{array} \right.
\end{eqnarray}
Figure \ref{Fig2_4}(b) shows the evolution of these two
dimensionless restoring forces for $w>0$.

\section{Theoretical resolution}\label{Section3}
In this paper, imperfections with dimensionless lengths $l<\sqrt{2}\pi$ are considered. In order to give a practical meaning to this inequation, let's consider the case of an imperfect railway.
The most common rails in France (rail type 50E6) have a weight per meter of about $m=500$ N m$^{-1}$, a length from $10$ to $400$ m (length between two joints connection) and a bending stiffness of about $EI=4\times10^6$ N m$^2$. Considering a coefficient of friction $\varphi$ between steel and concrete of $0.4$ and a mobilization $\Delta$ from $0.1$ to $1$ mm it comes a linear stiffness $K=\frac{m g \varphi}{\Delta}$ ($g$ being the gravity) from $2\times10^5$ to $2\times10^6$ N m$^{-2}$ and a characteristic length $L_{char}$ from $1$ to $2$ m. Therefore $l<\sqrt{2}\pi$  would correspond to a repeating sinusoidal imperfection in the railway whose half-wavelength is no more than $10$ m. In a more general way, $l<\sqrt{2}\pi$ deals with imperfections whose length does not exceed some meters. Therefore the specific calculations described in this paper have been made in the context of a small-scale experimental setup.

For $l<\sqrt{2}\pi$, the first buckling mode predicted by the linear analysis
is excited when
\begin{equation}
\lambda=\lambda_c  = \left( {\frac{{\pi }}{l}} \right)^2  + \left(
{\frac{{\pi }}{l}} \right)^{ - 2},
\end{equation}
and it has the same shape as the imperfection.
In what follows we
will also introduce the Euler load $\lambda_e$
\begin{equation}
\lambda_e  = \left( {\frac{{\pi }}{l}} \right)^2,
\end{equation}
which is the buckling load of the first mode when the restoring
force equals 0.

Two theories are developed to solve the equilibrium problem. The
first one, named piecewise solution theory is an exact resolution of
the equilibrium problem when the bi-linear restoring force is
considered. The second theory is based on a Galerkin method: it
leads to an approximate resolution of the equilibrium problem by
considering equally the bi-linear or the exponential restoring
force. To initiate this method, the deflection shape is assumed to
be a sinusoid, as the imperfection. Explicit solutions of the
Galerkin equation are obtained without any assumptions.

\subsection{Piecewise solution theory}
The principle of the piecewise solution theory is to solve
(\ref{chap4_eq11}) on each piece of the bi-linear function. Then,
the solutions are connected thanks to the boundary conditions and
assuming the continuity of $w$, $w^{'}$, $w^{''}$ and $w^{'''}$ at
two connecting points $x_1$ and $x_2$.

Substituting $p\left(w\right)$ by $-w$ and by $-1$ in
\eqref{chap4_eq11} yields respectively

\begin{subequations}
\begin{eqnarray}
{{w_1^{''''}}} + \lambda {{ w_1^{''}}} + w_1&=&-\lambda{ w_0^{''}},\label{EquationPourW1}\\
{{w_{2}^{''''}}} + \lambda {{ w_2^{''}}} &=&-1-\lambda{
w_0^{''}}.\label{EquationPourW2}
\end{eqnarray}
\end{subequations}
The solutions $w_1$ and $w_2$ belong to two affine spaces of
dimension 4 given by

\begin{subequations}
\begin{eqnarray}
w_1&=&{w_1}_{\rm{h}}+{{w_1}_{\rm{part}}},\label{SolutionW1}\\
w_2&=&{w_2}_{\rm{h}}+{{w_2}_{\rm{part}}},\label{SolutionW2}
\end{eqnarray}
\end{subequations}
where ${w_1}_{\rm{h}}$ and ${w_2}_{\rm{h}}$ satisfy the homogeneous
equations

\begin{subequations}\label{EquationHomogene}
\begin{eqnarray}
{{{{w_1}_{\rm{h}}}^{''''}}} + \lambda {{ {{w_1}_{\rm{h}}}^{''}}} +
{w_1}_{\rm{h}}&=&0,\label{EquationHomogeneW1}\\
{{{{w_2}_{\rm{h}}}^{''''}}} + \lambda {{ {{w_2}_{\rm{h}}}^{''}}}
&=&0.\label{EquationHomogeneW2}
\end{eqnarray}
\end{subequations}

Inserting ${w_1}_{\rm{h}}=A {\rm{e}}^{r_1x}$ and ${w_2}_{\rm{h}}=B
{\rm{e}}^{r_2x}$ in (\ref{EquationHomogene}) yields two algebraic
equations

\begin{subequations}
\begin{eqnarray}
r_{1}^4+\lambda {r_1}^2+1&=&0,\\
r_{2}^4+\lambda {r_2}^2&=&0,
\end{eqnarray}
\end{subequations}
whose solutions are

\begin{subequations}
\begin{eqnarray}
{r_1}^{\left(1\right)}&=&
 \frac{1}{2}\left[ { - 2\lambda  +2\left( {\lambda ^2  - 4} \right)^{{1 \mathord{\left/
 {\vphantom {1 2}} \right.
 \kern-\nulldelimiterspace} 2}} } \right]^{{1 \mathord{\left/
 {\vphantom {1 2}} \right.
 \kern-\nulldelimiterspace} 2}},\\
{r_1}^{\left(2\right)}&=&
 -\frac{1}{2}\left[ { - 2\lambda  + 2\left( {\lambda ^2  - 4} \right)^{{1 \mathord{\left/
 {\vphantom {1 2}} \right.
 \kern-\nulldelimiterspace} 2}} } \right]^{{1 \mathord{\left/
 {\vphantom {1 2}} \right.
 \kern-\nulldelimiterspace} 2}},\\
 {r_1}^{\left(3\right)}&=&
 \frac{1}{2}\left[ { - 2\lambda  - 2\left( {\lambda ^2  - 4} \right)^{{1 \mathord{\left/
 {\vphantom {1 2}} \right.
 \kern-\nulldelimiterspace} 2}} } \right]^{{1 \mathord{\left/
 {\vphantom {1 2}} \right.
 \kern-\nulldelimiterspace} 2}},\\
 {r_1}^{\left(4\right)}&=&
 -\frac{1}{2}\left[ { - 2\lambda  -2\left( {\lambda ^2  - 4} \right)^{{1 \mathord{\left/
 {\vphantom {1 2}} \right.
 \kern-\nulldelimiterspace} 2}} } \right]^{{1 \mathord{\left/
 {\vphantom {1 2}} \right.
 \kern-\nulldelimiterspace} 2}},
\end{eqnarray}
\end{subequations}
and
\begin{equation}
r_2 = 0,\pm \sqrt \lambda {\rm{i}}.
\end{equation}
Let's introduce $\alpha_i$ and $\omega_i$ the real and imaginary
parts of ${r_1}^{\left(i\right)}$. For $\lambda<2$, the roots $r_1$
are complex numbers and the function ${{w_1}_{\rm{h}}}$ is

\begin{eqnarray}
{{w_1}_{\rm{h}}}=A_1 e^{\alpha _1 x} \cos \left( {\omega _1 x}
\right) + A_2 e^{\alpha _2 x} \cos \left( {\omega _2 x} \right) +
A_3 e^{\alpha _1 x} \sin \left( {\omega _1 x} \right)\nonumber\\ +
A_4 e^{\alpha _2 x} \sin \left( {\omega _2 x} \right),
\end{eqnarray}
with $A_1$, $A_2$, $A_3$ and $A_4$ four real constants. For
$\lambda=2$, the roots $r_1$ are double imaginary numbers
(${r_1}^{\left(1\right)}={r_1}^{\left(3\right)}$ and
${r_1}^{\left(2\right)}={r_1}^{\left(4\right)}$). The function
${{w_1}_{\rm{h}}}$ is
\begin{equation}
{{w_1}_{\rm{h}}}=\left( {A_1x + A_2} \right)\cos \left( x \right) +
\left( {A_3x + A_4} \right)\sin \left( x \right).
\end{equation}
For $\lambda>2$, the roots $r_1$ are imaginary numbers. The solution
${{w_1}_{\rm{h}}}$ is
\begin{equation}
{{w_1}_{\rm{h}}}=A_1\cos \left( {\omega _1 x} \right) + A_2 \cos
\left( {\omega _3 x} \right) + A_3  \sin \left( {\omega _1 x}
\right) + A_4  \sin \left( {\omega _3 x} \right).
\end{equation}

For $\lambda\neq0$ the function ${{w_2}_{\rm{h}}}$ is given by
\begin{equation}
{{w_2}_{\rm{h}}} = B_1  + B_2 x + B_3 \cos \left( {\sqrt \lambda x}
\right) + B_4 \sin \left( {\sqrt \lambda  x} \right),
\end{equation}
with $B_1$, $B_2$, $B_3$ and $B_4$ four real constants.\\

Since ${{w_1}_{\rm{h}}}$ and ${{w_2}_{\rm{h}}}$ depend on 4
undetermined constants, these functions will be noted
${{w_1}_{\rm{h}}}\left(x,A_1,A_2,A_3,A_4\right)$ and
${{w_2}_{\rm{h}}}\left(x,B_1,B_2,B_3,B_4\right)$.

The functions ${{w_1}_{\rm{part}}}$ and ${{w_2}_{\rm{part}}}$
appearing in (\ref{SolutionW1}) and (\ref{SolutionW2}) are
particular solutions of (\ref{EquationPourW1}) and
(\ref{EquationPourW2}), with $w_0$ given by
(\ref{EquationDefautAdim}). Searching ${{w_1}_{\rm{part}}}$ and
${{w_2}_{\rm{part}}}$ as ${{w_1}_{\rm{part}}}=Aw_0$ and
${{w_2}_{\rm{part}}}=Cx^2+Aw_0$, yields

\begin{subequations}
\begin{eqnarray}
{{w_1}_{\rm{part}}}  &=& \frac{{\lambda
}}{{\lambda_c - \lambda
}}w_0,\\
{{w_2}_{\rm{part}}}  &=&- \frac{1 }{\lambda }\frac{{x^2 }}{2} +
\frac{{\lambda  }}{{{\lambda_e}  - \lambda}}w_0.
\end{eqnarray}
\end{subequations}

Finally, the functions $w_1$ and $w_2$, defined by
(\ref{SolutionW1}) and (\ref{SolutionW2}), are

\begin{subequations}
\begin{eqnarray}
w_1  &=& {{w_1}_{\rm{h}}} \left( {x, A_1 ,A_2 ,A_3 ,A_4 } \right) +
\frac{{\lambda
 }}{{\lambda_c  - \lambda }}w_0,\label{equationW1}\\
w_2  &=& {{w_2}_{\rm{h}}}\left( {x,B_1 ,B_2 ,B_3 ,B_4 } \right) -
\frac{1 }{\lambda }\frac{{x^2 }}{2} + \frac{{\lambda  }}{{\lambda_e
- \lambda}}w_0.\label{equationW2}
\end{eqnarray}
\end{subequations}

In what follows, we will introduce the function $w_3$ defined as
\begin{equation}\label{equationW3}
w_3  = {{w_1}_{\rm{h}}} \left( {x, C_1 ,C_2 ,C_3 ,C_4 } \right) +
\frac{{\lambda }}{{\lambda_c  - \lambda }}w_0.
\end{equation}
This function is a solution of (\ref{EquationPourW1}), as $w_1$, but
with 4 different undetermined constants $C_1$, $C_2$, $C_3$ and
$C_4$.

The aim of the piecewise solution theory is to search for a
deflection $w$ solution of (\ref{chap4_eq1}) by connecting the
functions $w_1$, $w_2$ and $w_3$ at two unknown points $x_1$ and
$x_2$ such that

\begin{equation}\label{ChampDeDeplacement}
w = \left\{ \begin{array}{l}
 w_1\hspace{1.37cm}{\rm{ if}}\hspace{0.1cm}
x \in \left[ {0,x_1 } \right[, \\
 1\hspace{1.58cm}{\rm{ if}}\hspace{0.1cm}
x=x_1,\\
 w_2 \hspace{1.37cm}{\rm{ if}}\hspace{0.1cm}x \in \left] {x_1,x_2 } \right[, \\
 1\hspace{1.58cm}{\rm{ if}}\hspace{0.1cm}
x=x_2,\\
 w_3 \hspace{1.37cm}{\rm{ if}}\hspace{0.1cm}x \in \left] {x_2,l } \right]. \\
 \end{array} \right.
\end{equation}
With these notations, the boundary conditions (\ref{chap4_eq12}),
(\ref{chap4_eq13}), (\ref{chap4_eq14}) and (\ref{chap4_eq15})
rewrites

\begin{subequations}\label{chap4_eq2}
\begin{eqnarray}
w_1\left( 0 \right) &=& 0,\label{chap4_eq21}\\
{{ w_1^{''}}}\left( 0 \right) &=& 0,\label{chap4_eq22}\\
w_3\left( l \right) &=&0,\label{chap4_eq23}\\
{{ w_3^{''}}}\left( l \right) &=& 0.\label{chap4_eq24}
\end{eqnarray}
\end{subequations}

The continuity of the displacement, the tangent, the curvature and
the shear at $x_1$ yields

\begin{subequations}\label{chap4_eq3}
\begin{eqnarray}
w_1\left( x_1 \right) -w_2\left( x_1 \right)&=&0 ,\label{chap4_eq31}\\
{{ w_1^{'}}}\left( x_1 \right) - {{ w_2^{'}}}\left( x_1 \right)&=&0,\label{chap4_eq32}\\
{{ w_1^{''}}}\left( x_1 \right) - {{ w_2^{''}}}\left( x_1 \right)&=&0,\label{chap4_eq33}\\
{{ w_1^{'''}}}\left( x_1 \right) - {{ w_2^{'''}}}\left( x_1
\right)&=&0,\label{chap4_eq34}
\end{eqnarray}
\end{subequations}
and at $x_2$
\begin{subequations}\label{chap4_eq4}
\begin{eqnarray}
w_2\left( x_2 \right) -w_3\left( x_2 \right)&=&0 ,\label{chap4_eq41}\\
{{ w_2^{'}}}\left( x_2 \right) - {{ w_3^{'}}}\left( x_2 \right)&=&0,\label{chap4_eq42}\\
{{ w_2^{''}}}\left( x_2 \right) - {{ w_3^{''}}}\left( x_2 \right)&=&0,\label{chap4_eq43}\\
{{ w_2^{'''}}}\left( x_2 \right) - {{ w_3^{'''}}}\left( x_2
\right)&=&0.\label{chap4_eq44}
\end{eqnarray}
\end{subequations}

Equations (\ref{chap4_eq2}), (\ref{chap4_eq3}) and (\ref{chap4_eq4})
lead to a linear system with 12 equations and 12 unknowns (i.e. the
amplitudes $A_i$, $B_i$ and $C_i$). A matrix representation of this
system is

\begin{eqnarray}\label{EqMatricielleG}
{\bf{G}}\left( {x_1 ,x_2 } \right) {\bf{a}} = {\bf{b}}\left( {w_0,
x_1, x_2 } \right),
\end{eqnarray}
with ${\bf{G}}$ a 12 by 12 real matrix and ${\bf{a}}$ the vector of
the unknown amplitudes. The vector ${\bf{b}}$ contains the
particular solutions ${{w_1}_{\rm{part}}}$ and ${{w_2}_{\rm{part}}}$
which depend on $w_0$, $x_1$ and $x_2$.

If $\det \left( {\bf{G}} \right)\neq0$ then

\begin{eqnarray}\label{EquationDesAmplitudes}
{\bf{a}}\left( {w_0, x_1 ,x_2 } \right) =  {\bf{G}}^{ - 1} \left(
{x_1 ,x_2 } \right) {\bf{b}}_{\bf{}} \left( {w_0, x_1 ,x_2 }
\right).
\end{eqnarray}

Equation (\ref{EquationDesAmplitudes}) express the amplitudes as
functions of $x_1$ and $x_2$. These two connecting points are
obtained by solving the nonlinear system

\begin{subequations}\label{SystemeNonlineaire}
\begin{eqnarray}
f_1\left(w_0, x_1, x_2\right)&=&w_1\left( {x_1} \right) -
1=0,\\
f_2\left(w_0, x_1,x_2\right)&=&w_2\left( {x_2 } \right) - 1=0.
\end{eqnarray}
\end{subequations}
The numerical resolution of this system has been carried out with
Matlab, using the {\textit{fzero}} function. This function tries to
find a zero of (\ref{SystemeNonlineaire}) near $X_0$, $X_0$ being a
vector of length two. Depending on $\lambda$
(\ref{SystemeNonlineaire}) has 0 or several solutions. For small
$\lambda$, the restoring force $p$ remains linear, such that $w=w_1$
for any $x\in[0,l]$. Thus, (\ref{SystemeNonlineaire}) has no
solution. For high $\lambda$, the restoring force $p$ is nonlinear:
the existence and the uniqueness of a solution for
(\ref{SystemeNonlineaire}) is not trivial. In this paper, we only
search for a solution which satisfies $x_1\in[0,l/2]$,
$x_2\in[l/2,l]$ and $x_2=l-x_1$: the connecting points are symmetric
relative to $x=l/2$. This condition is specified adjusting the $X_0$
vector used by the {\textit{fzero}} function. Typically, $X_0=[l/2;
l/2]$ is a good candidate to easily find the symmetric connecting
points. Once the connecting points are calculated, we determine the
amplitudes $A_i$, $B_i$ and $C_i$ thanks to
(\ref{EquationDesAmplitudes}), the functions $w_1$, $w_2$ and $w_3$
thanks to (\ref{equationW1}), (\ref{equationW2}), (\ref{equationW3})
and finally the deflection $w$ via (\ref{ChampDeDeplacement}).

\subsection{Galerkin method}\label{subsectionGalerkinMethod}

The Galerkin procedure \cite[see][]{Fox87} may be seen as being
derived from (\ref{VariationEnergiePotentielle}) by assuming that
the modes which go to make up $w$ are given by

\begin{equation}\label{GalerkinFonctionPhi}
w = \sum\limits_{i = 1}^n {y_i \phi _i } ,
\end{equation}
where each $y_i$ is an undetermined amplitude of each shape function
$\phi _i$. Depending on the form of $w$ in
(\ref{GalerkinFonctionPhi}), we can perform periodic or localized
buckling analysis. For a very long imperfection in the line of a strut on a cubic
foundation, \cite{Whiting17} performed the latter, using the
functions predicted by the asymptotical analysis
\cite[see][]{Wadee97} as test functions. The amplitudes of each
shape function are determined numerically using a variable-order
variable-step Adams method. In this paper we do not search for
localized solutions: we use a unique test function which has the
same shape as the imperfection. It means that the deflection $w$ is
searched as

\begin{equation} \label{ZEqnNum315473}
w=y\sin \left(\frac{\pi }{l} x\right),
\end{equation}
with $y>0$.

Inserting $\delta w$ in the dimensionless form of
(\ref{VariationEnergiePotentielle}) gives

\begin{equation}\label{IntegraleWhiting}
\int\limits_0^l {\sin \left( {\frac{\pi }{l}x} \right)\left[
{w^{''''}  + \lambda \left( {w^{''}  + w_0 ^{''} } \right) - p\left(
w \right)} \right]{\rm{dx}} = 0}.
\end{equation}

Writing the restoring force as $p\left(w\right)=-w-N\left(w\right)$
yields a relation between the load $\lambda$ and the amplitude $y$

\begin{equation} \label{1.6}
\lambda =\frac{1}{a_{0} +y} \left[\lambda_{c}
 y+\frac{Q\left(y\right)}{\lambda_{e} } \right],
\end{equation}
with
\begin{equation} \label{1.7}
Q\left(y\right)=\frac{2}{l} \int_{0}^{l}\sin \left(\frac{\pi }{l}
x\right)N\left(y\sin \left(\frac{\pi }{l} x\right)\right){\rm{dx}}.
\end{equation}

The assumption $N\left(y\sin \left(\frac{\pi }{l}
x\right)\right)=N\left(y\right)\sin \left(\frac{\pi }{l} x\right)$
introduced by \cite{Maltby8} yields $N=Q$. Such an assumption will
not be introduced in this paper. However, we will see how this
assumption affects the result.

Note that since the integrand function in (\ref{1.7}) is
$l/2$-periodic, the $Q$ function does not change when the deflection
is searched as $w=y\sin \left(\frac{n\pi }{l} x\right)$. Then, when
the imperfection is $w_0=a_0\sin \left(\frac{n\pi }{l} x\right)$,
the equilibrium paths are still given by (\ref{1.6}) with $\lambda_e
= \left( {\frac{{n\pi }}{l}} \right)^2$ and
$\lambda_c=\lambda_e+\lambda_e^{-1}$.

In practice, the equilibrium paths predicted by the Galerkin method
are plotted using (\ref{1.6}), varying $y$ and evaluating $\lambda$.

\subsubsection{Bi-linear restoring force}

Considering the bi-linear restoring force, the $Q$ function rewrites

\begin{equation} \label{2.1}
Q\left(y\right)=\frac{2{\rm{H}}\left(y-1 \right)}{\pi }
\left\{y\left[\arcsin\left(\frac{1 }{y} \right)-\frac{\pi }{2}
\right]+\left(\frac{y^{2} -1 }{y^{2} } \right)^{\frac{1}{2} }
\right\},
\end{equation}
where ${\rm{H}}$ is the Heaviside function defined by
${\rm{H}}\left( {y - 1} \right)=0$ if $y<1$ and ${\rm{H}}\left( {y -
1 } \right)=1$ if $y\geq1$. The proof of this result is reported in
\ref{Chap4Appendix1}.

\subsubsection{Exponential restoring force}

Considering the exponential restoring force, the $Q$ function
rewrites

\begin{equation} \label{ZEqnNum178974}
Q\left( y \right) = 2\left[ {{\mathop{{\rm I}}\nolimits} _1 \left( y
\right) - {\mathop{{\rm L}}\nolimits} _1 \left( y \right)} \right] -
y,
\end{equation}
where ${\rm{I}}_1$ and ${\rm{L}}_1$ are respectively the modified
Bessel and Struve functions of parameter 1. The proof of this result
is reported in \ref{Chap4Appendix2}.

\section{Theoretical and numerical results}\label{Section4}
In this section, the equilibrium paths predicted by the piecewise
solution theory, the Galerkin procedure and a numerical resolution
of (\ref{chap4_eq1}) are compared. We also determine the influence
of the restoring force (bi-linear or exponential) on the shape of
the equilibrium paths.

\subsection{Piecewise solution theory}
\begin{figure}
\begin{center}
\includegraphics[width=1\textwidth]{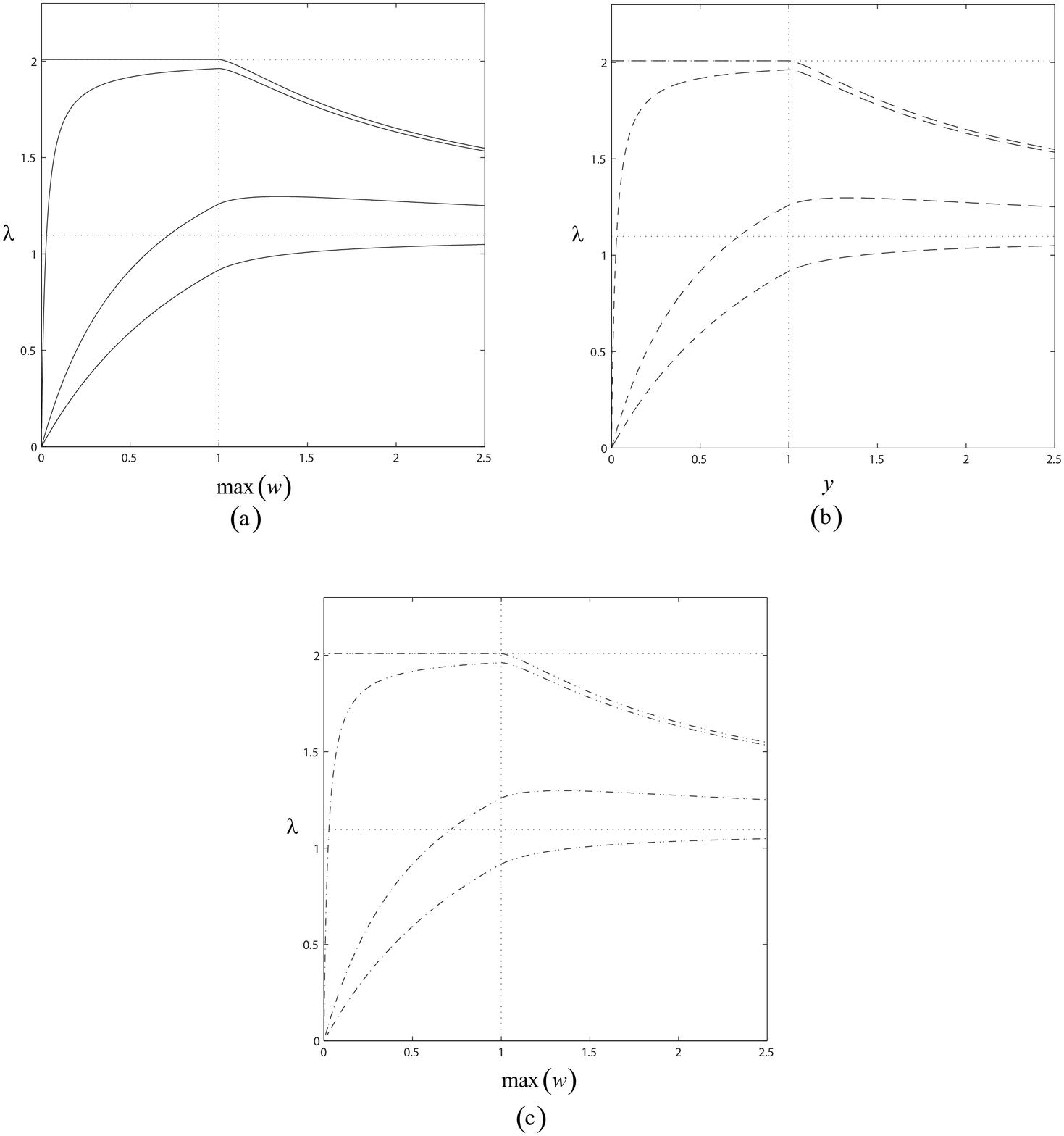}
\end{center}
\caption{Equilibrium paths predicted by (a) the piecewise solution
theory, (b) the Galerkin method, (c) the numerical resolution of
(\ref{chap4_eq1}), case of a bi-linear restoring force. On each
graph, the equilibrium paths are plotted (from top to bottom) for
$a_0=0$, $a_0=0.0238$, $a_0=0.595$, $a_0=1.19$ and $l=3$. Dotted
lines: critical load (upper line), Euler load (lower line),
mobilization (vertical line).}\label{ResultatsRaccordement1}
\end{figure}

\begin{figure}
\begin{center}
\includegraphics[width=1\textwidth]{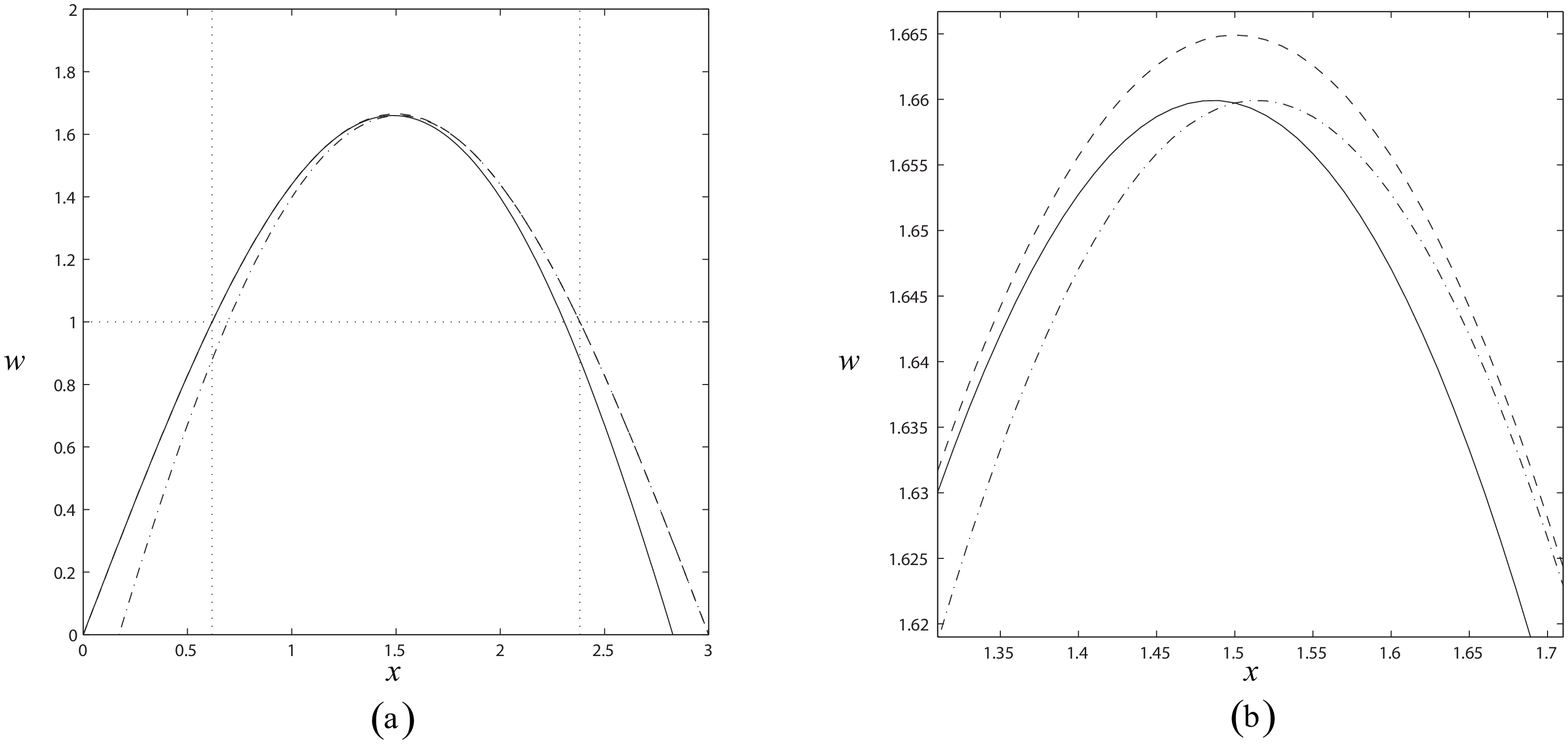}
\end{center}
\caption{Deflection predicted by the piecewise solution theory. (b)
is a zoom of (a). Function $w_1$ (solid line), function $w_2$
(dashed line), function $w_3$ (dash-dotted line). The functions
$w_1$ and $w_2$ are connected at $x=x_1\approx0.618$ (first vertical
dotted line). The functions $w_2$ and $w_3$ are connected at
$x=x_2\approx2.382$ (second vertical dotted line). The horizontal
dotted line represents the mobilization. $l=3$, $a_0=0$,
$\lambda=1.75$.}\label{ResultatsChampDeplacement}
\end{figure}

The equilibrium paths predicted by the piecewise solution theory are
depicted in figure \ref{ResultatsRaccordement1}(a) for $a_0$ from 0
to 1.19. For a small imperfection size, the equilibrium path shows
the load increasing at first but then hits a maximum (limit point,
or saddle-node bifurcation point) that is below $\lambda_c$ (or
equals for $a_0=0$), and the rest of the path asymptotically
decreases to the Euler load when
$\max\left(w\right)\rightarrow\infty$. Thus, the system is
subcritical. Moreover, the greater the imperfection size, the
greater the reduction in the maximum load. Thus the system is
imperfection sensitive. For high imperfection sizes, the equilibrium
path increases monotically and
$\lambda\rightarrow\lambda_e<\lambda_c$ when
$\max\left(w\right)\rightarrow\infty$.

From these observations we infer the existence of a critical
amplitude $a_{0c}$ such that

\begin{itemize}
    \item if $a_0>a_{0c}$ then the equilibrium paths do not have a limit point and the equilibrium states are stable,
    \item if $a_0<a_{0c}$ then the equilibrium paths have a limit point $\left(y_m,
    \lambda_m\right)$. For $y<y_m$ (resp. $y>y_m$)
    the equilibrium states are stable (resp. unstable). An unstable equilibrium state is represented in figure
\ref{ResultatsChampDeplacement} by connecting the functions $w_1$,
$w_2$, and $w_3$.
\end{itemize}
The determination of $a_{0c}$ is reported in section
\ref{sectionPointLimite}.

\subsection{Galerkin method and numerical results}

\subsubsection{Bi-linear restoring force}

For the bi-linear function, the equilibrium paths predicted by the
Galerkin method and those obtained via a numerical resolution of
(\ref{chap4_eq1}), using the ODE45 solver from Matlab (this routine
uses a variable step Runge-Kutta method), are represented in figures
\ref{ResultatsRaccordement1}(b) and \ref{ResultatsRaccordement1}(c).
The equilibrium paths are identical to those predicted by the
piecewise solution theory (the relative error between the two
theories and the numerical resolution being less than $0.1\%$).
Since the Galerkin test function and the imperfection have the same
shape, in the case of a bi-linear restoring force the deflection is
an amplification of the imperfection.

\subsubsection{Exponential restoring force}
\begin{figure}
\begin{center}
\includegraphics[width=1\textwidth]{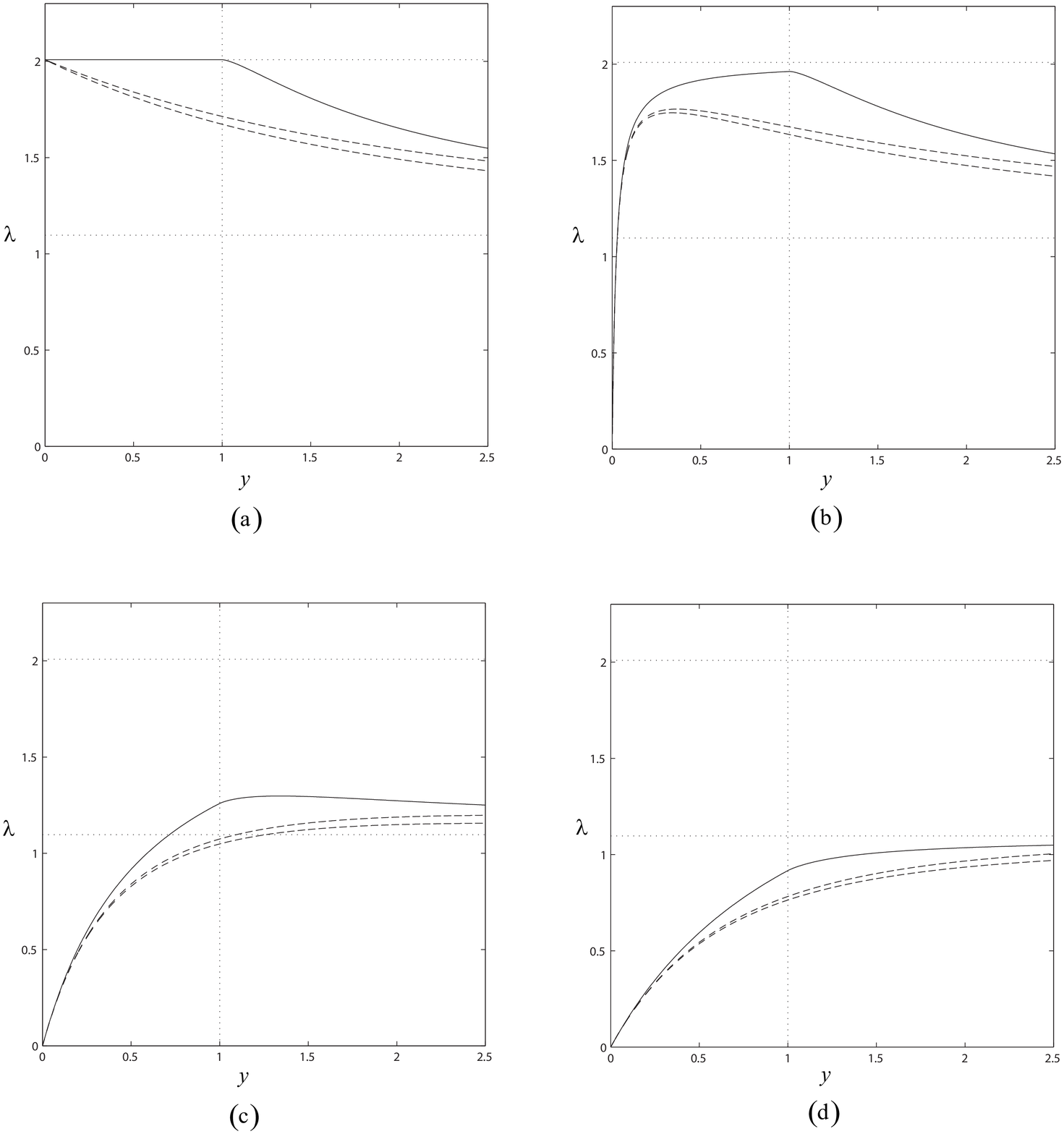}
\end{center}
\caption{Equilibrium paths predicted by the Galerkin method.
Bi-linear restoring force (solid line), exponential restoring force
(dashed lines). Galerkin method (upper dashed line), Galerkin method
with the \cite{Maltby8} assumption (lower dashed line). Dotted
lines: critical load (upper line), Euler load (lower line),
mobilization (vertical line). $l=3$, $a_0=0$ (a), $a_0=0.0238$ (b),
$a_0=0.595$ (c), $a_0=1.19$ (d).}\label{ResultatsTheoriques2}
\end{figure}

\begin{figure}
\begin{center}
\includegraphics[width=1\textwidth]{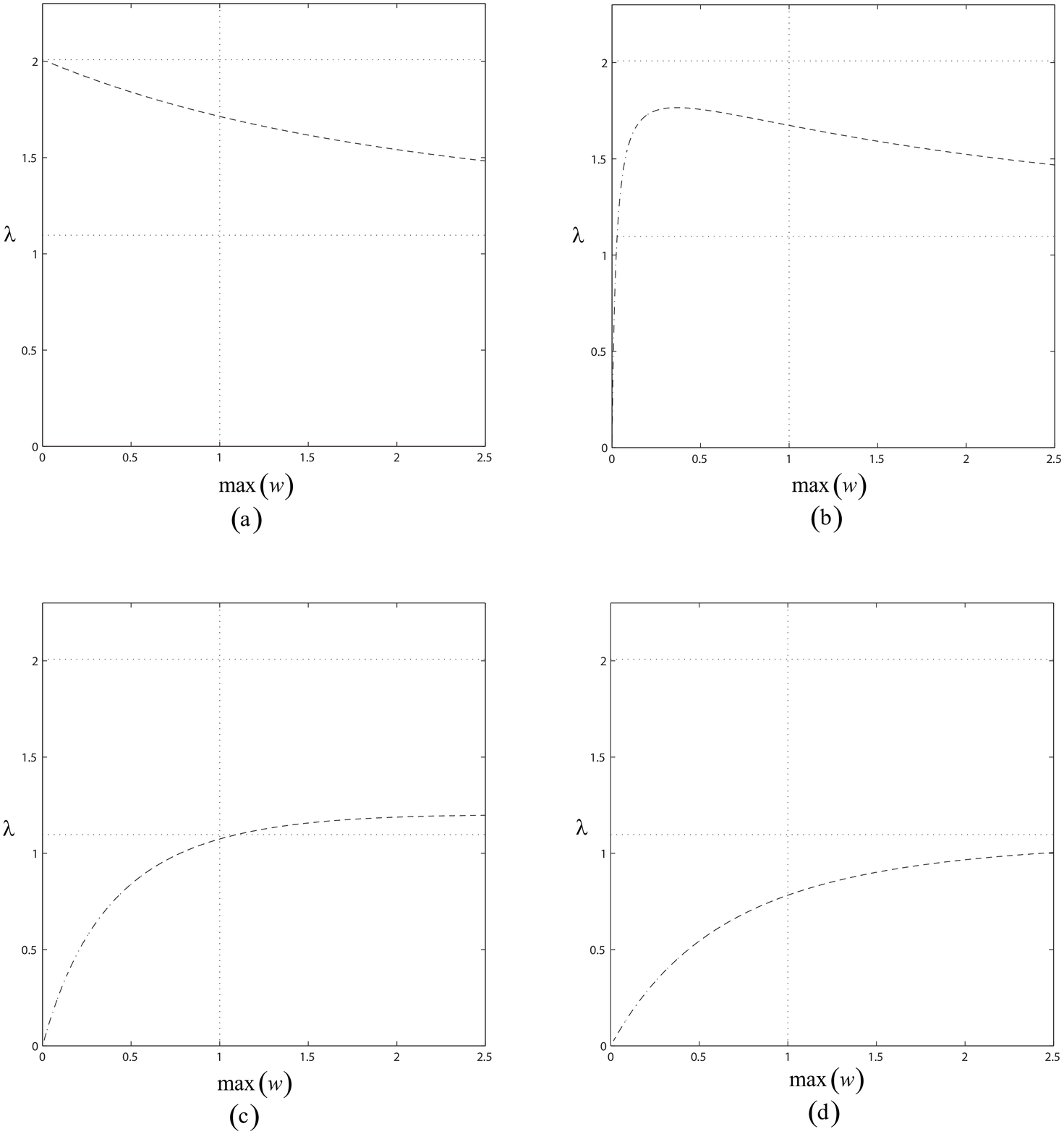}
\end{center}
\caption{Equilibrium paths obtained via a numerical resolution of
(\ref{chap4_eq1}), case of an exponential restoring force. Dotted
lines: critical load (upper line), Euler load (lower line),
mobilization (vertical line). $l=3$, $a_0=0$ (a), $a_0=0.0238$ (b),
$a_0=0.595$ (c), $a_0=1.19$ (d).}\label{ResultatsNumeriques2}
\end{figure}

For the exponential profile, the equilibrium paths predicted by the
Galerkin method and those obtained via a numerical resolution of
(\ref{chap4_eq1}), are represented in figures
\ref{ResultatsTheoriques2} and \ref{ResultatsNumeriques2}. Once
again they are identical, thus the deflection $w$ is an
amplification of the imperfection. The equilibrium paths predicted
by \cite{Maltby8} have also been reported in figure
\ref{ResultatsTheoriques2}. The assumption introduced by
\cite{Maltby8} (see section \ref{subsectionGalerkinMethod}) leads to
an underestimation of the load. In section \ref{sectionPointLimite}
we will show that the existence of a limit point is also affected by
this assumption.

In figure \ref{ResultatsTheoriques2} the equilibrium paths predicted
for a bi-linear restoring force have been reported in order to
discuss the influence of the restoring force. It appears that the
restoring force (bi-linear or exponential) has no influence on the
shape of the equilibrium paths (the variations are preserved, it
always exists a limit point for small imperfection sizes, the same
asymptotes are recovered). Nevertheless, the equilibrium paths for
an exponential profile are below the equilibrium paths for a
bi-linear profile. Therefore, the choice of the restoring force has
a non negligible influence on the maximal load acceptable by the
system. This point is the purpose of the next section.

\section{Limit point}\label{sectionPointLimite}
\begin{figure}
\begin{center}
\includegraphics[width=1\textwidth]{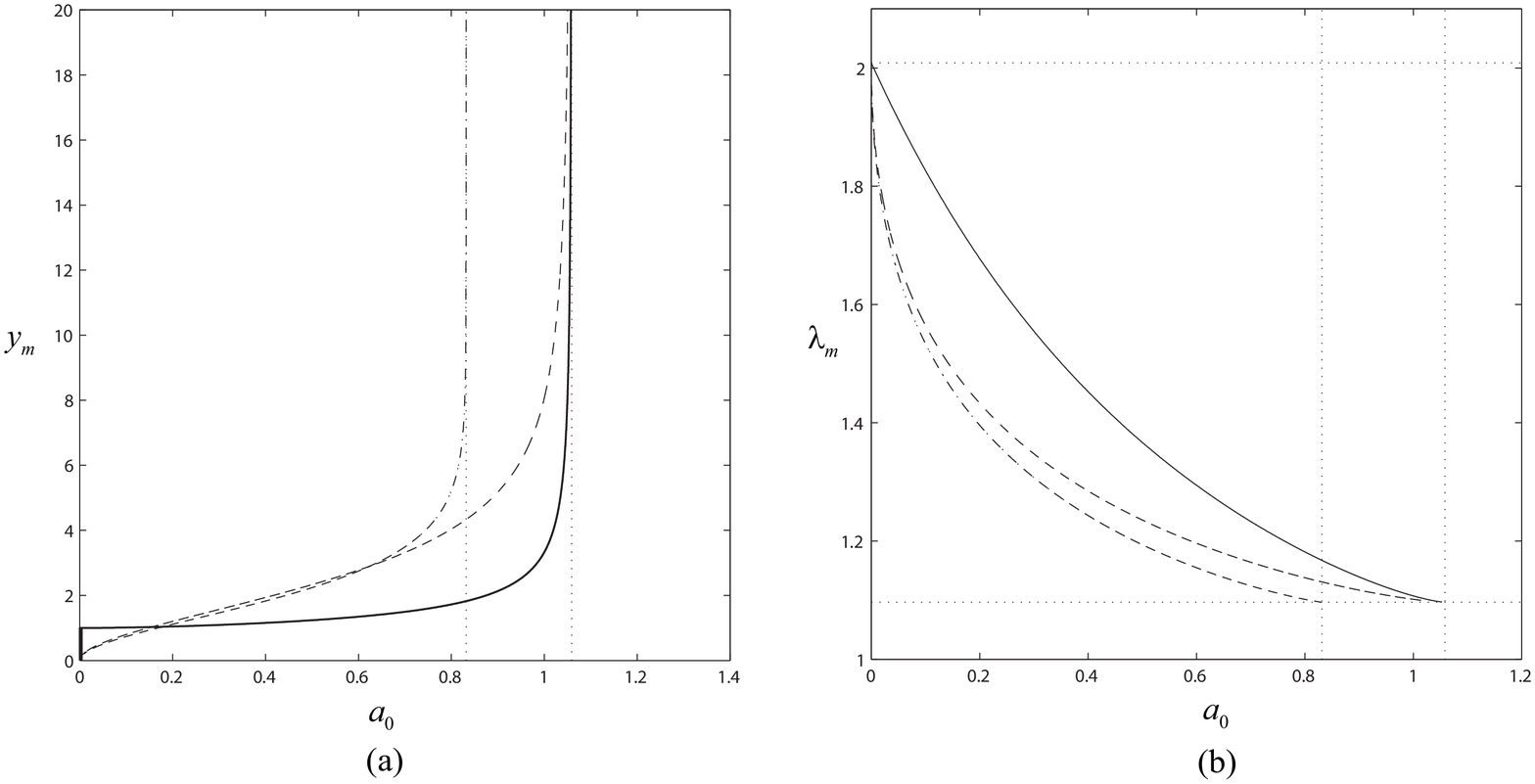}
\end{center}
\caption{Limit point as a function of the amplitude of the
imperfection. Bi-linear restoring force (solid line), exponential
restoring force (dashed line), exponential restoring force with the
assumption introduced by \cite{Maltby8} (dash-dotted line). Dotted
lines: (a) critical amplitudes, (b) critical load (upper line),
Euler load (lower line), critical amplitudes (vertical lines).
l=3.}\label{PointLimiteAdim}
\end{figure}

\begin{figure}
\begin{center}
\includegraphics[width=0.5\textwidth]{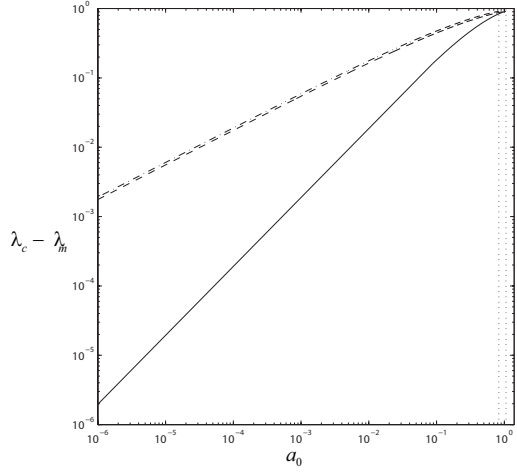}
\end{center}
\caption{Scaling of the limit point. Bi-linear restoring force
(solid line), exponential restoring force (dashed line), exponential
restoring force with the assumption introduced by \cite{Maltby8}
(dash-dotted line), critical amplitudes (dotted lines).
l=3.}\label{ScalingPointLimiteAdim}
\end{figure}

\begin{figure}
\begin{center}
\includegraphics[width=0.5\textwidth]{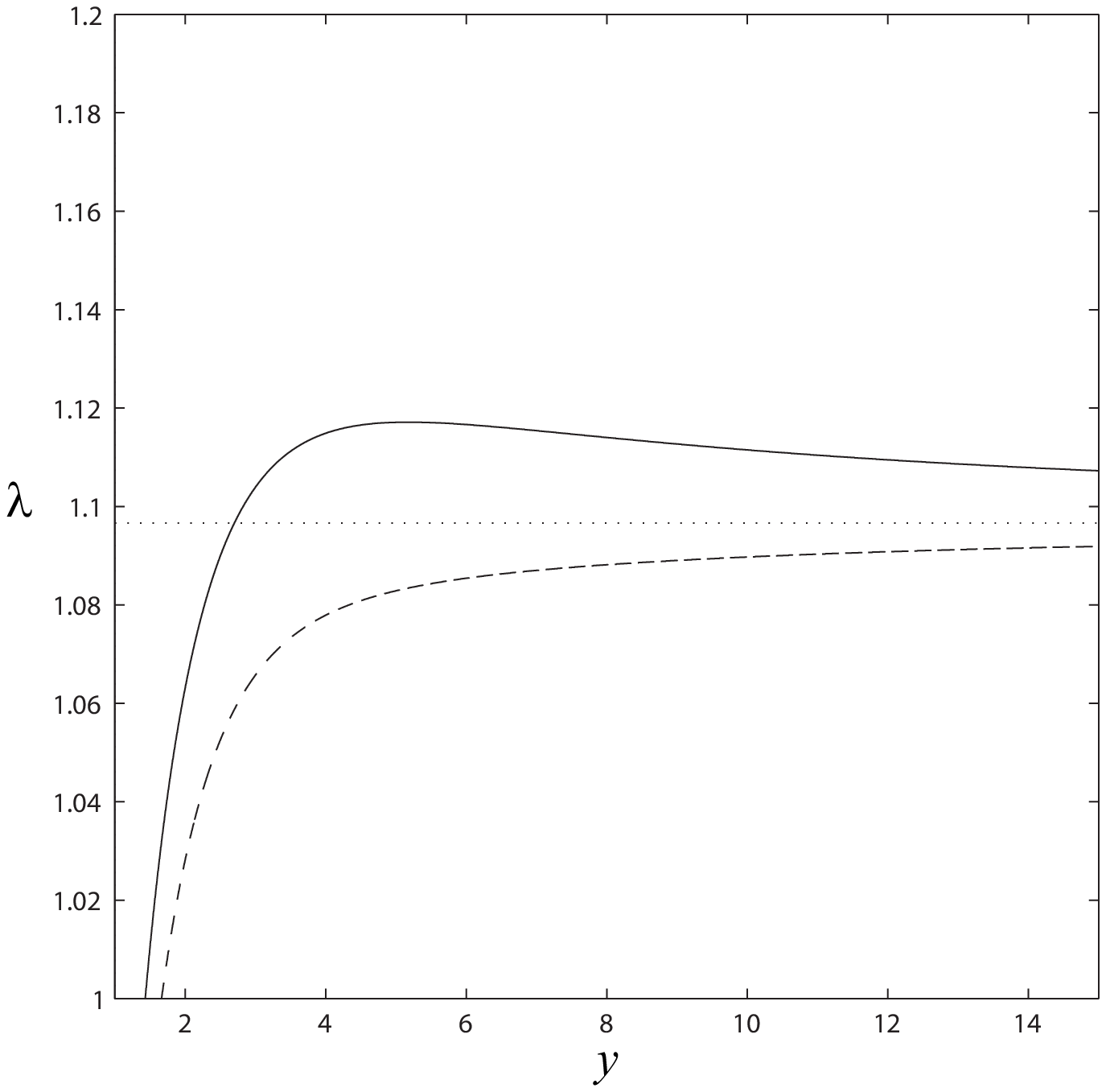}
\end{center}
\caption{Equilibrium paths predicted by the exact Galerkin method
(solid curve) and the Galerkin method with the assumption introduced
by \cite{Maltby8} (dashed line). The dotted line corresponds to the
Euler load. $a_0=0.9$, $l=3$.}\label{DerniereFigurePostDoc}
\end{figure}

A limit point corresponds to a maximum of $\lambda$. Therefore, if
it exists, this point $\left(y_m,\lambda_m\right)$ satisfies
$\frac{{d\lambda }}{{dy}}\left(y_m\right) = 0$, that is to say using
(\ref{1.6})

\begin{equation}\label{equationPointLimite}
\left( {a_0  + y_m } \right)\frac{{dQ}}{{dy}}\left( {y_m } \right) -
Q\left( {y_m } \right) =  - a_0 \lambda _c \lambda _e.
\end{equation}
This equation yields $\frac{{Q\left( {y_m } \right)}}{{\lambda _e }}
= a_0 \lambda _c - \left( {a_0  + y_m } \right){\mathop{{\rm
f}}\nolimits} \left( {a_0 } \right)$ with ${\rm{f}}$ a function of
$a_0$. Inserting this last expression in (\ref{1.6}) yields

\begin{equation}
\lambda _m  =\lambda\left(y_m\right)= \lambda _c  -
{\rm{f}}\left(a_0\right).\label{LambdaLimite}
\end{equation}

In practice, the coordinates of a limit point are obtained by
varying $y_m$ and looking for an amplitude $a_0$ which satisfies
(\ref{equationPointLimite}). Figure \ref{PointLimiteAdim}(a)
represents the evolution of $y_m$ as a function of $a_0$. For the
two restoring forces it appears that $y_m$ is an increasing function
of $a_0$ and diverges for the same critical amplitude. This
amplitude is determined by enforcing $y_m\rightarrow\infty$ in
(\ref{equationPointLimite}). It yields

\begin{equation}\label{AmplitudeCritique}
a_{0c}=\frac{4}{\pi}\lambda_e^{-2},
\end{equation}
whose dimensional equivalent form is

\begin{equation}
A_{0c}  = \frac{4}{\pi }\frac{{L^4 }}{{\pi ^4 EI}}K\Delta.
\end{equation}
This result is of great importance since it governs the appearance
of localized solutions, as demonstrated in \cite{Tveergard14}. In
that paper, they indeed showed the close link between the existence
of a limit point in the $\lambda\left(y\right)$ curve and
localization, the bifurcation into a localized form occurring a
little way beyond this limit point.

Considering the bi-linear restoring force, when $a_0\rightarrow0$,
equation (\ref{equationPointLimite}) has an infinity of solutions
$y_m$ in $\left] {0,1} \right]$. This observation is coherent with
the equilibrium paths depicted in figures
\ref{ResultatsRaccordement1} since the points lying on the plateau
$\lambda=\lambda_c$ are a maximum.

Considering the exponential restoring force, $y_m\rightarrow0$ when
$a_0\rightarrow0$. This result is coherent with the equilibrium
paths depicted in figures \ref{ResultatsTheoriques2}(a) and
\ref{ResultatsNumeriques2}(a) since $(0, \lambda_c)$ is a maximum.

Substituting $Q\left(y_m\right)$ by
$N\left(y_m\right)=-y_m+1-e^{-y_m}$ in (\ref{equationPointLimite})
(assumption made by \cite{Maltby8}) and enforcing
$y_m\rightarrow\infty$ yields
\begin{equation}
a_{0c}=\lambda_e^{-2},
\end{equation}
whose dimensional equivalent form is

\begin{equation}
A_{0c}  = \frac{{L^4 }}{{\pi ^4 EI}}K\Delta,
\end{equation}
which is $\frac{4}{\pi}$ times smaller than the exact prediction
(\ref{AmplitudeCritique}). This result means that a stable
equilibrium (i.e. an equilibrium path without a limit point)
predicted with the assumption introduced by \cite{Maltby8}, could in
fact be unstable (see figure \ref{DerniereFigurePostDoc}).

Figure \ref{PointLimiteAdim}(b) represents the evolution of the
limit point $\lambda_m$ as a function of $a_0$. Considering the two
restoring forces, the same asymptotes are observed:
$\lambda_m\rightarrow\lambda_e$ when $a_0\rightarrow a_{0c}$ and
$\lambda_m\rightarrow\lambda_c$ when $a_0\rightarrow0$. We also
observe that the assumption introduced by \cite{Maltby8} (see
section \ref{subsectionGalerkinMethod}) leads to an underestimation
of the limit load. These observations are coherent with the
equilibrium paths depicted in figures \ref{ResultatsRaccordement1},
\ref{ResultatsTheoriques2} and \ref{ResultatsNumeriques2}.

Figure \ref{ScalingPointLimiteAdim} shows the evolution of
$\lambda_c-\lambda_m$ as a function of $a_0$, in a logarithmic
scale. This picture allows to determine a scaling for the function
${\rm{f}}$ appearing in (\ref{LambdaLimite}). Considering the
bi-linear restoring force, for small amplitudes $a_0$,
$\lambda_m-\lambda_c$ scales as

\begin{equation}
\lambda _m -\lambda _c\sim - a_0.
\end{equation}
When the exponential restoring force is considered this scaling
becomes

\begin{equation}
\lambda _m  - \lambda _c\sim  - a_0 ^{{1 \mathord{\left/
 {\vphantom {1 2}} \right.
 \kern-\nulldelimiterspace} 2}}.
\end{equation}
We observe in figure \ref{ScalingPointLimiteAdim} that the
assumption introduced by \cite{Maltby8} has no influence on this
last scaling.

\section{Conclusion}

In this paper, the growth of a repeating sinusoidal imperfection in the line of a strut on a nonlinear elastic Winkler type foundation is considered. The imperfection is introduced by considering an initially sinusoidal
deformed shape with an half wavelength. The imperfection length is chosen
such that the buckle mode predicted by the linear theory has the
same shape as the imperfection (first buckle mode). The
nonlinearities are only due to the restoring force provided by the
foundation. This restoring force is expressed as a
force-displacement relationship which is either a bi-linear or an
exponential function. The equilibrium problem is solved using three
different methods. The first one, named piecewise solution theory,
is dedicated to the bi-linear profile and leads to an exact
resolution of the equilibrium problem. The second one is available
whatever the restoring force and is based on a Galerkin procedure.
This procedure is initiated with a test function which has the same
shape as the imperfection. It yields an explicit relation between
the compressive load and the amplitude of the test function. This
expression is an exact solution of the Galerkin equation and gives
an approximate solution of the equilibrium problem. The last method
is a numerical resolution of the equilibrium problem, using the
ODE45 solver from Matlab. These three solving methods yield the same
results: whatever the restoring force (bi-linear or exponential),
the bifurcation is subcritical, the system is imperfection sensitive
and the deformed shape is an amplification of the default. Moreover,
it exists a critical imperfection size
$a_{0c}=\frac{4}{\pi}\lambda_e^{-2}$ ($\lambda_e$ being the Euler
load) which does not depend on the restoring force and such that

\begin{itemize}
    \item if $a_0>a_{0c}$, then the equilibrium path shows the load increasing
monotically and remains asymptotic to the Euler load.
    \item if $a_0<a_{0c}$, then the equilibrium path shows the load increasing
at first but then hits a limit point and the rest of the path is
asymptotic to the Euler load.
\end{itemize}
This paper provides a better estimate of $a_{0c}$ with respect to
previous publications.

For each restoring fore, an approximate mathematical rule is derived
relating the imperfection size $a_0$ to the corresponding limit load
$\lambda_m$. Considering the bi-linear profile (resp. the
exponential profile) the limit point scales as
$\lambda_m-\lambda_c\sim -a_0$ (resp. $\lambda_m-\lambda_c\sim
-a_0^{1/2}$), where $\lambda_c$ is the critical load issued from the
classical linear analysis. Therefore, the scaling of the limit point
depends on the regularization method.

In this paper, the restoring force and the compressive load are
independent. Nevertheless, in some industrial applications (such as
in drilling problems) the restoring force slightly depends on the
axial compressive load. Therefore, we are currently carrying out a
study with a bi-linear restoring force proportional to the axial
load. First results issued from the Galerkin approach indicate that
it is necessary to redefine the dimensionless parameters, leading to
new scalings for the critical imperfection size and the limit load.

\appendix


\section{Function $Q$, bi-linear restoring force}\label{Chap4Appendix1}
The aim of this Appendix is to calculate the function $Q$ appearing
in (\ref{2.1}) when the bi-linear restoring force is considered.

Introduce ${\rm{H}}$ the Heaviside function defined as
${\rm{H}}\left(x\right)=0$ if $x<0$ and ${\rm{H}}\left(x\right)=1$
if $x\geq0$. The restoring force $p$ rewrites

\begin{eqnarray}\label{AnnexeP}
p =  -w - \left( {{\rm{Sgn}}\left(w\right) - w }
\right){\rm{H}}\left( {\left| w \right|
 - 1 } \right),
\end{eqnarray}
where ${\rm{Sgn}}$ is the sign function. Since
$p=-w-N\left(w\right)$ it comes

\begin{eqnarray}\label{AnnexeN}
N\left(w\right) =  \left( {{\rm{Sgn}}\left(w\right) - w }
\right){\rm{H}}\left( {\left| w \right|
 - 1 } \right).
\end{eqnarray}
For an imperfection with an half-wavelength, it can be assumed that
$w>0$. Then, the $Q$ function is (see (\ref{1.7}))

\begin{equation}
Q\left(y\right)=-\frac{2}{l} \int_{0}^{l}\left[y\sin \left(\frac{\pi
}{l} x\right)-1 \right]\sin \left(\frac{\pi }{l}
x\right){\rm{H}}\left( {y\sin \left(\frac{\pi }{l} x\right)
 - 1 } \right){\rm{dx}}.
\end{equation}

If $y<1$ then ${y\sin \left( {\frac{\pi }{l}x} \right) - 1 }<0$ so
${\rm{H}}\left( {y\sin \left( {\frac{\pi }{l}x} \right) - 1 }
\right)=Q=0$. Therefore, the function $Q$ can be written as

\begin{eqnarray}
Q \left( y \right) =-
\frac{2{\rm{H}}\left(y-1\right)}{l}\int\limits_0^l {\left[ {y\sin
\left( {\frac{\pi }{l}x} \right) - 1 } \right]} \sin \left(
{\frac{\pi }{l}x} \right){\rm{H}}\left( {y\sin \left( {\frac{\pi
}{l}x} \right) - 1 } \right){\rm{dx}}.
\end{eqnarray}
The change of variable $t=\frac{\pi}{l}x$ gives

\begin{eqnarray}
Q \left( y \right) =-
\frac{2}{\pi}{\rm{H}}\left(y-1\right)\int\limits_0^{\pi} {\left[
{y\sin \left( t \right) - 1 } \right]} \sin \left(
t\right){\rm{H}}\left( {y\sin \left( t \right) - 1 }
\right){\rm{dt}}.
\end{eqnarray}
The argument of the Heaviside function under the integral sign
equals 0 when $\sin\left(t\right)=\frac{1}{y}$, that is to say for
$t=t_1=\arcsin\left(\frac{1}{y}\right)$ and $t={\pi}-t_1$. It comes
that the function $Q$ is non zero between $t_1$ and ${\pi}-t_1$

\begin{eqnarray}
Q \left( y \right) =-
\frac{2}{\pi}{\rm{H}}\left(y-1\right)\int\limits_{t_1}^{{\pi}-t_1}
{\left[ {y\sin \left( t \right) - 1 } \right]} \sin \left(
t\right){\rm{dt}}.
\end{eqnarray}
Finally, this integral yields
\begin{eqnarray}
Q \left( y \right) = \frac{2}{\pi }{\rm{H}}\left( {y - 1 }
\right)\left\{ {y\left[\arcsin\left( {\frac{1 }{y}} \right)
-\frac{\pi }{2}\right ] +  \left( {\frac{{y^2  - 1 }}{{y^2 }}}
\right)^{\frac{1}{2}} } \right\},
\end{eqnarray}
and the result from (\ref{2.1}) is recovered.


\section{Function $Q$, exponential restoring force}\label{Chap4Appendix2}
The aim of this Appendix is to calculate the function $Q$ appearing
in (\ref{ZEqnNum178974}) when the exponential restoring force is
considered. For this calculus, we recall that the modified Struve
and Bessel functions of parameter 1 can be expended as power series
\begin{subequations}\label{SeriesStruveBessel}
\begin{eqnarray}
{\rm{L}} _1 \left( y \right) &=&  \sum\limits_{p = 1}^{ + \infty }
{\frac{2}{\pi }\frac{{\left( {p!} \right)^2 }}{{\left( {2p + 1}
\right)\left[ {\left( {2p} \right)!} \right]^2 }}\left( {2y}
\right)^{2p} },\\
{\rm{I}}_1 \left( y \right) &=& \sum\limits_{p = 0}^{ + \infty }
{\frac{1}{{p!\left( {p + 1} \right)!}}\left( {\frac{y}{2}}
\right)^{2p + 1} }.
\end{eqnarray}
\end{subequations}

Substituting the exponential function in
(\ref{LoiRessortRegularise}) by its power series yields

\begin{equation}
p\left(w\right)=-w+ \sum _{n=2}^{+\infty }\frac{\left(-1\right)^{n}
w^{n} }{ n!},
\end{equation}
so that

\begin{equation}
N\left(w\right)=\sum _{n=2}^{+\infty }\frac{\left(-1\right)^{n+1}
w^{n} }{ n!}.
\end{equation}
Therefore, (\ref{1.7}) gives

\begin{equation}
Q\left(y\right)=\frac{2 }{l} \int _{0}^{l}\sum _{n=2}^{+\infty
}\frac{\left(-1\right)^{n+1} \sin \left(\frac{\pi }{l}
x\right)^{n+1} }{ n!} y^{n}  {\rm{dx}}.
\end{equation}
Inverting the sum and integral signs and introducing the change of
variable $t=\frac{\pi }{l} x$ yields

\begin{equation}\label{AnnexeSerieQ}
Q\left(y\right)=\frac{4 }{\pi } \sum _{n=2}^{+\infty
}\frac{\left(-1\right)^{n+1} W_{n+1} }{ n!} y^{n},
\end{equation}
with $W_{n} $ the Wallis integral

\begin{equation}
W_n  = \int\limits_0^{{\pi  \mathord{\left/
 {\vphantom {\pi  2}} \right.
 \kern-\nulldelimiterspace} 2}} {\sin ^n \left( x \right){\rm{dx}}}.
\end{equation}

The terms $W_n$ are classical to calculate. For $n=2p$ and $n=2p+1$
it yields

\begin{subequations}
\begin{eqnarray}\label{AnnexeWallisTermes}
 W_{2p}  &=& \frac{{\left( {2p} \right)!}}{{2^{2p} \left( {p!} \right)^2 }}\frac{\pi
 }{2},\\
 W_{2p + 1}  &=& \frac{{2^{2p} \left( {p!} \right)^2 }}{{\left( {2p +
1}
 \right)!}}.
\end{eqnarray}
\end{subequations}

Splitting the serie appearing in (\ref{AnnexeSerieQ}) into odd and
even indices gives $ Q\left( y \right) = \Sigma _1 \left( y \right)
+ \Sigma _2 \left( y \right) $ with

\begin{subequations}
\begin{eqnarray}\label{AnnexeSigma1et2}
\Sigma _1 \left( y \right) &=&  - \frac{4}{\pi }\sum\limits_{p =
1}^{
+ \infty } {\frac{{W_{2p + 1} }}{{\left( {2p} \right)!}}y^{2p} },\\
\Sigma _2 \left( y \right) &=& \frac{4}{\pi }\sum\limits_{p = 1}^{ +
\infty } {\frac{{W_{2\left( {p + 1} \right)} }}{{\left( {2p + 1}
\right)!}}y^{2p + 1} }.
\end{eqnarray}
\end{subequations}

Inserting (\ref{AnnexeWallisTermes}) in (\ref{AnnexeSigma1et2})
yields

\begin{subequations}
\begin{eqnarray}
\Sigma _1 \left( y \right) &=&  - 2\sum\limits_{p = 1}^{ + \infty }
{\frac{2}{\pi }\frac{{\left( {p!} \right)^2 }}{{\left( {2p + 1}
\right)\left[ {\left( {2p} \right)!} \right]^2 }}\left( {2y}
\right)^{2p} },\\
\Sigma _2 \left( y \right) &=& 2\sum\limits_{p = 0}^{ + \infty }
{\frac{1}{{p!\left( {p + 1} \right)!}}\left( {\frac{y}{2}}
\right)^{2p + 1} }  - y.
\end{eqnarray}
\end{subequations}
Equation (\ref{SeriesStruveBessel}) gives

\begin{subequations}
\begin{eqnarray}
\Sigma _1 \left( y \right) &=&  - 2{\mathop{{\rm L}}\nolimits}_1 \left( y \right),\\
\Sigma _2 \left( y \right) &=& 2{\mathop{{\rm I}}\nolimits} _1
\left( y \right) - y,
\end{eqnarray}
\end{subequations}
with respectively ${\mathop{{\rm L}}\nolimits}_1$ and ${\mathop{{\rm
I}}\nolimits} _1$ the modified Struve and Bessel functions of
parameter 1. Finally
\begin{equation}
Q\left( y \right) = \Sigma _1 \left( y \right) + \Sigma _2 \left( y
\right) = 2\left[ {{\mathop{{\rm I}}\nolimits} _1 \left( y \right) -
{\mathop{{\rm L}}\nolimits} _1 \left( y \right)} \right] - y,
\end{equation}
and the result from (\ref{ZEqnNum178974}) is recovered.


\bibliographystyle{elsart-harv}
\bibliography{Biblio}

\end{document}